\documentclass[journal]{IEEEtran}
\usepackage{latexsym}
\usepackage{amsfonts,amssymb,amsmath}
\usepackage{hyperref}
\def\BibTeX{{\rm B\kern-.05em{\sc i\kern-.025em b}\kern-.08em T\kern-.1667em\lower.7ex\hbox{E}\kern-.125emX}}
\usepackage{subfigure}
\usepackage{algorithm}
\usepackage{algpseudocode}
\usepackage{amsmath}
\usepackage{amsfonts}
\usepackage{graphicx}
\usepackage{epsfig}
\usepackage{cite}
\usepackage{txfonts}
\usepackage{relsize}
\usepackage{color}
\usepackage{color}
\usepackage{CJK}
\usepackage{times}
\usepackage{multicol}
\usepackage{stfloats}
\usepackage{float}
\usepackage{lipsum}
\usepackage{bm}

\begin{document}
\title{On the performance of Active STAR-RIS-Assisted Cell-Free Massive MIMO Systems with Phase Errors and Channel Aging
\thanks{This work was supported by the Research Grants Council under the Area of Excellence scheme grant AoE/E-601/22-R.}}

\author{Jun~Qian,~\IEEEmembership{Member,~IEEE,}
        Ross~Murch,~\IEEEmembership{Fellow,~IEEE,}
and~Khaled~B.~Letaief,~\IEEEmembership{Fellow,~IEEE}
\thanks{The authors are with the Department of Electronic and Computer Engineering, The Hong Kong University of Science and Technology,
Hong Kong (e-mail: eejunqian@ust.hk, eermurch@ust.hk, eekhaled@ust.hk).}}

\maketitle
%\IEEEtitleabstractindextext
{\begin{abstract}
  Active reconfigurable intelligent surfaces (RISs) employ amplification to overcome attenuation caused by the RIS cascaded link. In this paper, we analyze the effects of phase errors and channel aging in active simultaneously transmitting and reflecting (STAR) RIS-assisted cell-free massive multiple-input multiple-output (MIMO) systems. By leveraging a spatially correlated Rayleigh fading model, this paper derives minimum mean square error estimate-based channel estimates and formulates closed-form expressions for downlink spectral efficiency. This analytical framework enables a comprehensive evaluation of the effects of channel aging and uniformly distributed phase errors on system performance. The results demonstrate that active STAR-RISs can effectively compensate for the adverse effects of phase errors and channel aging. To counteract the impact of channel aging, we propose practical guidelines for resource-block-length design. Also, an increase in APs and STAR-RIS elements, along with a larger amplification factor, can alleviate performance degradation.

\end{abstract}

% Note that keywords are not normally used for peerreview papers.
}

\maketitle

\vspace{-6pt}\section{Introduction}
To satisfy the ever-increasing demand for wireless communication, various advanced technologies are being considered \cite{9570143,7827017}. Among these, cell-free massive multiple-input multiple-output (MIMO) has been developed to mitigate inter-cell interference and expand transmission coverage by integrating massive MIMO with distributed networks \cite{7827017,10297571,qian2024impactchannelagingelectromagnetic}. Reconfigurable intelligent surfaces (RISs) have also been proposed to enhance spectral efficiency (SE) by providing additional controllable and reconfigurable cascaded links \cite{10297571}.

Recently, integrating RISs with cell-free massive MIMO shows the potential for synergistic benefits, particularly beneficial with unreliable access point (AP)-user links 
\cite{9875036,9665300,10640072}. However, RIS-assisted networks function best when the receivers and transmitters are on the same side of RISs\cite{10297571,qian2024impactchannelagingelectromagnetic}, while realistic users are placed on both RIS sides\cite{10297571,10373089}. Thus, \cite{9570143} developed simultaneously transmitting and reflecting (STAR) RISs, supporting receivers and transmitters on both RIS sides. 
Besides passive STAR-RISs, active STAR-RISs have gained attention for their ability to amplify signals incident on them to compensate for the cascaded AP-RIS-user channels \cite{10264149,10643432}. This amplification potentially allows active STAR-RIS to outperform passive STAR-RIS \cite{10264149,10643432}.

Integrating STAR-RISs with cell-free massive MIMO has been suggested to combine the advantages of STAR-RISs and cell-free massive MIMO systems\cite{10297571,11196010,10841966}. \cite{10297571} studied the implications of spatial correlation in STAR-RIS-assisted cell-free massive MIMO systems. \cite{10841966} analyzed the hardware impairments, and \cite{11196010} analyzed the effects of electromagnetic interference and phase errors on these systems. However, limited research has explored active STAR-RIS-assisted cell-free massive MIMO, with only \cite{10264149} assessing the correlated active STAR-RIS-assisted cell-free massive MIMO systems.

One vital issue in RIS and STAR-RIS-assisted networks is the phase errors caused by imperfect phase estimation in the RIS phase shifting network \cite{10373089,9786058,10841966}. \cite{10373089} and \cite{9786058} studied the sum-rate bound and the coverage probability of phase error-affected STAR-RIS-assisted networks. \cite{10841966} analyzed STAR-RIS-assisted cell-free massive MIMO systems experiencing hardware impairments and phase errors. Compounding these issues, user mobility introduces channel aging, resulting in Doppler shifts and outdated channel state information (CSI)\cite{9875036,9416909}. This poses challenges in STAR-RIS-assisted cell-free massive MIMO systems, as mobile users are common in real-world applications\cite{9875036,qian2024impactchannelagingelectromagnetic,9416909}.

To our knowledge, no comprehensive study has investigated the combined effects of phase errors and channel aging on active STAR-RIS-assisted cell-free massive MIMO systems.
However, as these practical factors can impair performance, an analysis is required to investigate the performance limits and system design guidelines. Therefore, we address this research gap in this paper, and our contributions include:

\begin{itemize}
    \item  We establish an active STAR-RIS-assisted cell-free massive MIMO model with phase errors and channel aging, including spatially correlated Rayleigh fading channels.

    \item We derive closed-form expressions for the downlink SE that explicitly account for phase errors and channel aging to analyze the system performance. 

    \item Results show that increasing the number of APs and STAR-RIS elements effectively alleviates performance degradation caused by phase errors and channel aging. A resource-block-length design guideline to counteract channel aging is proposed. The results also show that active STAR-RISs can improve performance compared to passive STAR-RISs with phase errors and channel aging.

\end{itemize}

\vspace{-11pt}\section{System Model}
 
The active STAR-RIS-assisted cell-free massive MIMO system is modelled with $K$ single-antenna users simultaneously served by $M$ APs, equipped with $N$ antennas, linked to the central processing unit (CPU) through perfect fronthaul links\cite{10468556,11196010}. An active $L$-element STAR-RIS facilitates communication between users and APs. Users in the STAR-RIS reflection and transmission areas are indexed by sets $\mathcal{K}_r$ and $\mathcal{K}_t$, respectively. The cardinalities of the sets are $|\mathcal{K}_r|=K_r$ and $|\mathcal{K}_t|=K_t$ where $K_r+K_t=K$ and $\mathcal{K}_r\cap\mathcal{K}_t=\varnothing$. To define the STAR-RIS operation mode, $\omega_k=r$ denotes the operation mode for the $k$-th reflection-area user. Similarly, $\omega_k=t$ denotes the operation mode for the $k$-th transmission-area user \cite{10297571}.

\vspace{-12pt}\subsection{STAR-RIS-assisted Channel Model with Phase Errors}
This paper focuses on the system operating under the time-division duplex (TDD) mode over spatially correlated Rayleigh fading channels\cite{9416909}. The aggregate uplink channel from the $k$-th user to the $m$-th AP, facilitated by the active STAR-RIS, at the $n$-th time instant, can be obtained by\cite{qian2024impactchannelagingelectromagnetic}
 \vspace{-3pt}
\begin{equation}
     \displaystyle
     \textbf{g}_{mk}[n]=\underbrace{\displaystyle \sqrt{\beta_{mk}}{\textbf{R}}_{m,r}^{1/2}\textbf{v}_{mk}[n]}_{\textbf{d}_{mk}[n]}+\textbf{g}_{m}[n]\boldsymbol{\Theta}_{\omega_k}[n]\textbf{g}_{k}[n],
    \label{cascaded_uplink_channel_via_RIS}
   \end{equation}
   where $\textbf{d}_{mk}[n]\in \mathbb{C}^{N\times 1}$ is the direct link from the $k$-th user to the $m$-th AP at the $n$-th time instant. $\beta_{mk}$ represents the large-scale fading coefficient between the $m$-th AP and the $k$-th user. ${\textbf{R}}_{m,r} \in \mathbb{C}^{N\times N}$ represents the $m$-th AP spatial correlation matrix. The independent fast-fading channel, $\textbf{v}_{mk}[n]\sim \mathcal{CN}(\textbf{0},\textbf{I}_{N})$, contains independent and identically distributed (i.i.d.) random variables distributed as $\mathcal{CN}(0,1)$. Following the classical Kronecker channel model in \cite{qian2024impactchannelagingelectromagnetic}, the channel, $\textbf{g}_{m}[n]\in \mathbb{C}^{N\times L}$, from the active STAR-RIS to the $m$-th AP, can be given by
    \vspace{-3pt}
\begin{equation}
     \displaystyle \textbf{g}_{m}[n]=\sqrt{\beta_{m}}{\textbf{R}}_{m,r}^{1/2}\textbf{v}_{m}[n]{\textbf{R}}_{m,t}^{{1/2}},
     \label{channel_RIS_AP}
   \end{equation}
where $\beta_{m}$ denotes the large-scale fading coefficient between the $m$-th AP and the STAR-RIS. ${\textbf{R}}_{m,r}  \in \mathbb{C}^{N\times N}$ is the spatial correlation matrix of the $m$-th AP, $\textbf{v}_{m}[n]\in \mathbb{C}^{N\times L}$ comprises i.i.d. random variables, sampled from $\mathcal{CN}(0,1)$. ${\textbf{R}}_{m,t}=A{\textbf{R}} \in \mathbb{C}^{L\times L}$ denotes the STAR-RIS spatial correlation matrix. With the vertical height $d_V$ and the horizontal width $d_H$, $A=d_Vd_H$ represents the area of a STAR-RIS element\cite{9300189}. The $(m,n)$-th element in $\textbf{R}$ is formulated as $ [\textbf{R}]_{x,y}=\text{sinc}\Big{(}{2||\textbf{u}_x-\textbf{u}_y||}/{\lambda_c}\Big{)}$ \cite{9300189},
in which $\text{sinc}(a)=\text{sin}(\pi a)/(\pi a)$ is the sinc function and $\lambda_c$ represents the carrier wavelength. The position vector for the $x$-th STAR-RIS element is $\textbf{u}_x=[0,\text{mod}(x-1,L_h)d_h,\lfloor(x-1)/L_h\rfloor d_v]^T$\cite{9300189}, with $L=L_h\times L_v$, where the numbers of column and row elements are $L_h$ and $L_v$, respectively.

Since APs and STAR-RIS are static, constant $\textbf{g}_{m}[n]=\textbf{g}_{m}$, $\forall n$, are assumed within the resource block. Then, phase shift matrices are pre-determined with $\boldsymbol{\Theta}_{\omega_k}[n]= \boldsymbol{\Theta}_{\omega_k}=\textbf{A}\bar{\boldsymbol{\Phi}}_{\omega_k}{\boldsymbol{\Phi}}_{\omega_k}$, $n=1,...,\tau_c$. $\textbf{A}=\text{diag}\left(\sqrt{\alpha_1},...,\sqrt{\alpha_L}\right)$ is the active STAR-RIS's amplification factor matrix \cite{9786058}. The commonly used energy splitting protocol is adopted herein\cite{11196010,10297571}, where the STAR-RIS behavior is governed by the coefficient matrices $\boldsymbol{\Phi}^{}_{r}=\text{diag}(u_{1}^r{\phi}_{1}^r,u_{2}^r{\phi}_{2}^r,...,u_{L}^r{\phi}_{L}^r)\in\mathbb{C}^{L\times L}$ for the reflection mode and $\boldsymbol{\Phi}_{t}=\text{diag}(u_{1}^t{\phi}_{1}^t,u_{2}^t{\phi}_{2}^t,...,u_{L}^t{\phi}_{L}^t)\in\mathbb{C}^{L\times L}$ for the transmission mode. The terms ${\phi}_{l}^t=e^{i\theta_{l}^t},~{\phi}_{l}^r=e^{i\theta_{l}^r}$ represent the induced phase shifts with $\theta_{l}^t,~\theta_{l}^r \in[0,2\pi)$, and $u_{l}^t,~u_{l}^r \in[0,1]$, with $(u_{l}^t)^2+(u_{l}^r)^2=1,~\forall l$, are amplitude coefficients \cite{9570143,10264149}. The phase error matrix is $\bar{\boldsymbol{\Phi}}_{\omega_k}=\text{diag}(e^{i\bar{\theta}_{1}^{\omega_k}},e^{i\bar{\theta}_{2}^{\omega_k}},...,e^{i\bar{\theta}_{L}^{\omega_k}})\in\mathbb{C}^{L\times L}$ \cite{10025392,9786058}. Based on \cite{10025392,10373089}, phase errors can be modelled by i.i.d. random zero-mean variables, with the uniformly distributed $\bar{\theta}_{l}^{\omega_k}\in[-\kappa,\kappa], ~\forall l$, having the characteristic function $\mathbb{E}\{e^{i\bar{\theta}_{l}^{\omega_k}}\}=\frac{\text{sin}(\kappa)}{\kappa}=\phi,~\forall l$\cite{10025392}. 
Note that other practical factors, such as mutual coupling between closely-spaced elements \cite{10440504} and external electromagnetic interference impinging on STAR-RISs\cite{9598875,11196010}, also greatly impact performance. Due to space constraints, a comprehensive analysis of these factors will be addressed in future work.
Then, the channel from the $k$-th user to the STAR-RIS, $\textbf{g}_{k}[n]\in \mathbb{C}^{L\times 1}$, is formulated as
    \vspace{-3pt}
     \begin{equation}
     \begin{array}{c@{\quad}c}
\textbf{g}_{k}[n]=\sqrt{\beta_{k}}{\textbf{R}}_{k}^{1/2}\textbf{v}_{k}[n].
     \end{array}
     \label{channel_user_RIS}
   \end{equation}
where $\beta_{k}$ denotes the large-scale fading coefficient between the $k$-th user and the STAR-RIS. The STAR-RIS spatial correlation matrix is ${\textbf{R}}_{k} =A{\textbf{R}} \in \mathbb{C}^{L\times L}$\cite{9300189}. The fast-fading channel is defined as $\textbf{v}_{k}[n]\in \mathbb{C}^{L\times 1}$, composed of i.i.d. random variables following $\mathcal{CN}(0,1)$. For ease of explanation, the covariance matrix of the aggregate channel vector $\textbf{g}_{mk}[n]$ is formulated as
    \vspace{-5pt}
\begin{equation}
     \begin{array}{ll}
\displaystyle \mathbf{\Delta}_{mk} = \mathbb{E}\{\textbf{g}_{mk}[n]\textbf{g}_{mk}[n]^H\}\vspace{3 pt}
=\beta_{mk}\textbf{R}_{m,r}+\beta_{m}\beta_{k}\textbf{R}_{m,r}\text{tr}({\textbf{T}}_{\omega_k}),
     \end{array}
     \setlength{\belowdisplayskip}{1pt}
\label{RIS_spatial_correlation_element}    \vspace{-5pt}
   \end{equation}
   with ${\textbf{T}}_{\omega_k} \displaystyle ={\textbf{R}}_{m,t}^{1/2}\boldsymbol{\Theta}_{\omega_k}\textbf{R}_{k}\boldsymbol{\Theta}_{\omega_k}^H{\textbf{R}}_{m,t}^{1/2}\displaystyle=A^2{\textbf{R}}^{1/2}\bar{\textbf{R}}_{\omega_k}{\textbf{R}}^{1/2}$ and $\bar{\textbf{R}}_{\omega_k}=\mathbb{E}\{\boldsymbol{\Theta}_{\omega_k}\textbf{R}\boldsymbol{\Theta}_{\omega_k}^H\}=\textbf{A}\boldsymbol{\Phi}_{\omega_k}\left(\phi^2\textbf{R}+(1-\phi^2)\textbf{R}\circ\textbf{I}_L\right)\boldsymbol{\Phi}_{\omega_k}^H\textbf{A}^H.$

\vspace{-11pt}\subsection{Channel Aging}
Due to user mobility, the channel aging effect occurs \cite{9875036}. Thus, the flat-fading channel coefficients keep constant within each symbol and vary by symbol \cite{9875036}. Based on \cite{qian2024impactchannelagingelectromagnetic,10468556}, we model the aggregate uplink channel $\textbf{g}_{mk}[n]$ as the sum of initial states $\textbf{g}_{mk}[0]$ and innovation components
    \vspace{-3pt}
\begin{equation}
     \displaystyle \textbf{g}_{mk}[n]=\rho_k[n]\textbf{g}_{mk}[0]+\bar{\rho}_k[n]\textbf{e}_{mk}[n],
     \label{uplink_channel_channel_aging}
   \end{equation}
where $\rho_k[n]$, with $\bar{\rho}_k[n]=\sqrt{1-{\rho}_k^2[n]}$, is the $k$-th user's temporal correlation coefficient between channel realizations at time instants $0$ and $n$. The independent innovation component at time instant $n$ \cite{9416909} is formulated as $\textbf{e}_{mk}[n]=\textbf{e}_{mk}^\text{d}[n]+\textbf{g}_{m}\boldsymbol{\Theta}_{\omega_k}\textbf{e}_{k}[n]\sim\mathcal{CN}(\textbf{0},\mathbf{\Delta}_{mk}))$. 

\textit{Remark 1}: Given that ${\rho}_k[n]=J_0(2\pi f_{D,k}T_sn)$ \cite{9416909,qian2024impactchannelagingelectromagnetic}, where $J_0(\cdot)$ is the zeroth-order Bessel function of the first kind, the Doppler shift for the $k$-th user is $f_{D,k}=\frac{v_kf_c}{c}$ with user velocity $v_k$, carrier frequency $f_c$, and the speed of light $c=3\times 10^8~ $m/s. $T_s$ is the time instant length.

\vspace{-11pt}
\section{Uplink Channel Estimation and Downlink SE Analysis}
This section introduces the active STAR-RIS-assisted uplink channel estimation with pilot symbols\cite{10297571,11196010}. The subsequent derivation of novel closed-form expressions for downlink SE analyzes the downlink system performance.

\vspace{-11pt}\subsection{Uplink Channel Estimation}
During channel estimation, users send pre-allocated pilots to APs. Subsequently, the APs acquire the uplink CSI by employing the minimum mean square error (MMSE) estimate\cite{10468556}. $\varphi_k \in \mathbb{C}^{\tau_p\times 1}$ with $\varphi_k^{H}\varphi_k=1,\forall k$ represents the time-multiplexed pilot assigned to the $k$-th user. typically, the resource block length $\tau_c$ is significantly larger than the pilot sequence length $\tau_p$\cite{9875036}. Note that $K>\tau_p$, the same pilot sequence will be assigned to multiple users, leading to pilot contamination \cite{qian2024impactchannelagingelectromagnetic,10468556}.
The $t$-th pilot sequence is transmitted only at the $t$-th time instant within the resource block \cite{9416909,qian2024impactchannelagingelectromagnetic,10468556}. The time instant allocated to $k$-th user can be indexed by $t_k \in\{1,...,\tau_p\}$ and those users sharing the same pilot sequence as the $k$-th user follow $\mathcal{P}_k=\{k'~:~t_{k'}=t_k\}\subset\{1,...,K\}$. Inevitably, $\varphi_k^H \varphi_{k'}=1, \forall k^{\prime} \in \mathcal{P}_k$. %Inevitably, $\varphi_k^{H}\varphi_{k'}=1, \forall k' \in \mathcal{P}_k$. 
We express
the pilot sequence received at the $m$-th AP, $\textbf{Y}_{m,p}[t_k]\in \mathbb{C}^{N\times\tau_p}$, as\cite{9786058,qian2024impactchannelagingelectromagnetic}
\vspace{-5 pt}
\begin{equation}
\begin{array}{ll}
     \displaystyle \textbf{Y}_{m,p}[t_k]&      \displaystyle =\sum\nolimits_{k'=1}^{K}\sqrt{p_p}\textbf{g}_{mk'}[t_{k'}]\varphi_{k'}^{H}\\
     & \displaystyle+\textbf{g}_{m}\boldsymbol{\Theta}_{{r}}\textbf{V}_{r}[t_{k}]+\textbf{g}_{m}\boldsymbol{\Theta}_{t}\textbf{V}_{t}[t_{k}]+\textbf{W}_{m,p}[t_k],
\end{array}\label{received_pilot_signal_jth_phase}
   \end{equation}
where $\textbf{V}_{\omega}[t_{k}]\in\mathbb{C}^{L\times\tau_p}$ is inevitable dynamic noise introduced by active
STAR-RIS elements where the $v$-th column satisfies $\Big{[}\textbf{V}_{\omega}[t_{k}]\Big{]}_v\sim \mathcal{CN}(0,\sigma_v^2\textbf{I}_{L})$. The static noise is negligible and omitted compared to the dynamic noise\cite{9786058}. $\textbf{W}_{m,p}[t_k]\in\mathbb{C}^{N\times\tau_p}$ is the additive white Gaussian noise (AWGN) matrix with the $v$-th column $\Big{[}\textbf{W}_{m,p}[t_k]\Big{]}_v\sim \mathcal{CN}(\textbf{0},\sigma^2\textbf{I}_{N})$. This work provides the channel estimates at the $\lambda$-th time instant with $\lambda=\tau_p+1$. {\color{black}Given the channel temporal correlations across different time instants in \eqref{uplink_channel_channel_aging}, the acquired estimates will be adopted as the initial
states to introduce channel estimates at the other time instants}\cite{9416909,qian2024impactchannelagingelectromagnetic}. The effective channel at time instant $t_{k'}$ is 
    \vspace{-3pt}
\begin{equation}
\begin{array}{ll}
\displaystyle \textbf{g}_{mk'}[t_{k'}]&\displaystyle=\rho_{k'}[\lambda-t_{k'}]\textbf{g}_{mk'}[\lambda]\vspace{3 pt}+\bar{\rho}_{k'}[\lambda-t_{k'}]\textbf{e}_{mk'}[t_{k'}].
\end{array}
   \end{equation}
    \vspace{-4pt}Therefore, the projection $\textbf{y}_{mk,p}[t_k]\displaystyle =\textbf{Y}_{m,p}[t_k]\varphi_k$ is given by
 \vspace{-2pt}
\begin{equation}
\begin{array}{ll}
     \displaystyle \textbf{y}_{mk,p}[t_k]
    & \displaystyle=\sum\nolimits_{{k'}\in\mathcal{P}_{k}}\sqrt{p_p}\textbf{g}_{mk'}[t_{k'}]+\textbf{g}_{m}\boldsymbol{\Theta}_{{r}}\textbf{V}_{r}[t_{k}]\varphi_k\\
    & \displaystyle+\textbf{g}_{m}\boldsymbol{\Theta}_{t}\textbf{V}_{t}[t_{k}]\varphi_k+\textbf{W}_{m,p}[t_k]\varphi_k. 
\end{array}\label{projected_pilot_signal_j-th_phase}
   \end{equation}
Based on \eqref{projected_pilot_signal_j-th_phase}, the MMSE of $\textbf{g}_{mk}[\lambda]$ is defined as
 \vspace{-5pt}
\begin{equation}
\begin{array}{ll}
     \displaystyle \hat{\textbf{g}}_{mk}[\lambda]=\sqrt{p_p}\rho_k[\lambda-t_k]\mathbf{\Delta}_{mk}\mathbf{\Psi}_{mk}^{-1}\textbf{y}_{mk,p}[t_k],
\end{array}\label{LMMSE_RIS_channel_estimation}
   \end{equation}
    \vspace{-14pt}
   \begin{equation}
 \setlength{\abovedisplayskip}{5pt}
\begin{array}{ll}
\displaystyle
\mathbf{\Psi}_{mk}&\displaystyle=p_p\sum\nolimits_{k'\in\mathcal{P}_k}\mathbf{\Delta}_{mk'}+\beta_m\sigma_v^2\textbf{R}_{m,r}\Big{(}\text{tr}(\mathbf{\Gamma}_t)+\text{tr}(\mathbf{\Gamma}_r)\Big{)}
+\sigma^2\textbf{I}_N,
\end{array}
\label{Psi}
   \end{equation}
where $\mathbf{\Gamma}_\omega=A\textbf{R}^{1/2}\boldsymbol{\Theta}_{\omega}\boldsymbol{\Theta}_{\omega}^H\textbf{R}^{1/2}$, $\omega=t,r$. Meanwhile, $ \hat{\textbf{g}}_{mk}\sim\mathcal{CN}\left(\textbf{0},\textbf{Q}_{mk}\right)$ with $\textbf{Q}_{mk}=p_p\rho_k[\lambda-t_k]^2\mathbf{\Delta}_{mk}\mathbf{\Psi}_{mk}^{-1}\mathbf{\Delta}_{mk}$.
  Therefore, the proposed channel model is formulated as \cite{9416909}
      \vspace{-3pt}
\begin{equation}
\begin{array}{ll}
     \displaystyle \textbf{g}_{mk}[n]
   \displaystyle =\displaystyle  \rho_k[n-\lambda]{\textbf{g}}_{mk}[\lambda]+
     \bar{\rho}_k[n-\lambda]\textbf{e}_{mk}[n].
\end{array}
\label{channel_estimate}
   \end{equation}

\vspace{-12pt}\subsection{Downlink SE Analysis}
The broadcast channel, where the precoding vector $\textbf{f}_{mk}[n]\in\mathbb{C}^{N\times 1}$, is applied to compose the downlink data transmission from all APs to all users \cite{9875036}. We model the downlink channel as the transpose of the uplink channel, utilizing the TDD channel reciprocity characteristic\cite{9416909,qian2024impactchannelagingelectromagnetic}. We express the transmit signal from the $m$-th AP at time instant $n$ as
\vspace{-5 pt}
\begin{equation}
\begin{array}{ll}
     \displaystyle \textbf{x}_m[n]=\sqrt{p_d}\sum\nolimits_{k=1}^K \textbf{f}_{mk}[n]\sqrt{\eta_{mk}}q_k[n],
\end{array}\label{transmitted_signal_m_AP}
   \end{equation}
where $p_d$ denotes the downlink transmit power, $q_k[n]\sim \mathcal{CN}(0,1)$ denotes the signal of the $k$-th user. $\eta_{mk},~\forall m,~\forall k,$ denotes the power control coefficient satisfying the power constraint $\mathbb{E}\big{\{}|\textbf{x}_m[n]|^2\big{\}}\leq p_d$. Here, we utilize conjugate beamforming $\textbf{f}_{mk}[n]=\hat{\textbf{g}}_{mk}^{\text{*}}[\lambda]$ for $\lambda\leq n\leq \tau_c$ \cite{qian2024impactchannelagingelectromagnetic}.
\eqref{downlink_received_signal_k_user} at the top of the next page formulates the
signal received at the $k$-th user at time instant $n$, in which $\textbf{v}[n]\sim\mathcal{CN}(0,\sigma_v^2\textbf{I}_L)$ denotes the dynamic noise at the active STAR-RIS, the receiver noise at the $k$-th user is $w_k[n]\sim \mathcal{CN}(0,\sigma^2)$.
\begin{figure*}[!t]
\begin{equation}
\begin{array}{ll}
     \displaystyle r_k[n]&   \displaystyle =\sum\nolimits_{m=1}^{M}\textbf{g}_{mk}^T[n]\textbf{x}_m[n]+\textbf{g}_{k}[n]^T\boldsymbol{\Theta}_{\omega_k}^T\textbf{v}[n]+w_k[n]\vspace{3 pt}\\&   \displaystyle 
     =\underbrace {\sqrt{p_d}\rho_k[n-\lambda]\sum\nolimits_{m=1}^{M}\mathbb{E}\Big{\{}\textbf{g}_{mk}^{T}[\lambda]\hat{\textbf{g}}_{mk}^*[\lambda]\Big{\}}\sqrt{\eta_{mk}}q_k[n]}_{\text{DS}_k[n]}\vspace{3 pt}+\underbrace {\sqrt{p_d}\rho_k[n-\lambda]\sum\nolimits_{m=1}^{M}\Big{(}\textbf{g}_{mk}^{T}[\lambda]\hat{\textbf{g}}_{mk}^*[\lambda]-\mathbb{E}\Big{\{}\textbf{g}_{mk}^{T}[\lambda]\hat{\textbf{g}}_{mk}^*[\lambda]\Big{\}}\Big{)}\sqrt{\eta_{mk}}q_k[n]}_{\text{BU}_k[n]}\\&\displaystyle+\underbrace {\sqrt{p_d}\bar{\rho}_k[n-\lambda]\sum\nolimits_{m=1}^{M}\textbf{e}_{mk}^{T}[n]\hat{\textbf{g}}_{mk}^*[\lambda]\sqrt{\eta_{mk}}q_k[n]}_{\text{CA}_k[n]}\vspace{3 pt}+\displaystyle\sum\nolimits_{k'\neq k}^K\underbrace {\sqrt{\rho_d} \sum\nolimits_{m=1}^{M}\textbf{g}_{mk}^{T}[n]\hat{\textbf{g}}_{mk'}^*[\lambda]\sqrt{\eta_{mk'}}q_{k'}[n]}_{\text{UI}_{kk'}[n]}+\underbrace{\textbf{g}_{k}[n]^T\boldsymbol{\Theta}_{\omega_k}^T\textbf{v}[n]}_{\text{DN}_k[n]}+\underbrace {w_{k}[n]}_{\text{NS}_k[n]},
\end{array}\label{downlink_received_signal_k_user}
\vspace{-8 pt}
   \end{equation}
    \vspace{-18 pt}
   \hrulefill
\end{figure*}
Then, we express the lower-bound downlink SE of the $k$-th user as \cite{9416909}
\vspace{-5pt}
\begin{equation}
\begin{array}{ll}
     \displaystyle \text{SE}_k=\frac{1}{\tau_c}\sum\nolimits_{n=\lambda}^{\tau_c}\text{log}_2\Big{(} 1+\text{SINR}_k[n]\Big{)},
\end{array}\label{downlink_SE_description}
   \end{equation}
where $\text{SINR}_k[n]$ stands for the effective signal-to-interference-plus-noise ratio (SINR) at time instant $n$. Given space limitations, we exclude the full proof of \eqref{downlink_received_signal_k_user}. However, it can be derived by following the computation in appendices of \cite{qian2024impactchannelagingelectromagnetic, 11196010}.
The summation of inter-user interference can be derived by \eqref{DL_downlink_UI} with $\textbf{Z}_{mk}=\sqrt{p_p}\rho_k[\lambda-t_k]\mathbf{\Delta}_{mk}\mathbf{\Psi}_{mk}^{-1},~\forall m,~\forall k$, at the top of the next page.
   \begin{figure*}[!t]
\begin{equation}
\begin{array}{ll}
\displaystyle\sum\nolimits_{k'=1}^K\mathbb{E}\{|\text{UI}_{kk'}[n]|^2\} \displaystyle=p_d\rho_k^2[n-\lambda]\sum\nolimits_{k'\in\mathcal{P}_k}\sum\nolimits_{m=1}^M\eta_{mk'}p_p\rho_k^2[\lambda-t_k]
(\beta_m\beta_k)^2\text{tr}\left(\textbf{T}_{\omega_k}^2\right)\text{tr}\left(\textbf{R}_{m,r}\textbf{Z}_{mk'}^H\textbf{R}_{m,r}\textbf{Z}_{mk'}\right)\\ ~~~~~~~~~~~~~~~~~~~\displaystyle+p_d\rho_k^2[n-\lambda]\sum\nolimits_{k'\in\mathcal{P}_k}\sum\limits_{m=1}^M\sum\limits_{n=1}^M\sqrt{\eta_{mk'}\eta_{nk'}}p_p\rho_k^2[\lambda-t_k]\text{tr}\Big{(}\mathbf{\Delta}_{mk}\textbf{Z}_{mk'}^H\Big{)}\text{tr}\Big{(}\mathbf{\Delta}_{nk}\textbf{Z}_{nk'}\Big{)}
\\ ~~~~~~~~~~~~~~~~~~~\displaystyle+p_d\sum\limits_{k'=1}^K\sum\limits_{m=1}^M\eta_{mk'}
\left[
p_p\sum\limits_{k''\in\mathcal{P}_{k'}}\text{tr}\left(\mathbf{\Delta}_{mk}\textbf{Z}_{mk'}^H\mathbf{\Delta}_{mk''}\textbf{Z}_{mk'}\right)+\beta_m\sigma_v^2\text{tr}\left(\mathbf{\Gamma}_r+\mathbf{\Gamma}_t\right)\text{tr}\left(\mathbf{\Delta}_{mk}\textbf{Z}_{mk'}^H\textbf{R}_{m,r}\textbf{Z}_{mk'}\right)+\sigma^2\text{tr}\left(\mathbf{\Delta}_{mk}\textbf{Z}_{mk'}\textbf{Z}_{mk'}^H\right)\right]
\\ ~~~~~~~~~~~~~~~~~~~
\displaystyle+p_d\sum\limits_{k'=1}^K\sum\limits_{m=1}^M\sum\limits_{n=1}^M\sqrt{\eta_{mk'}\eta_{nk'}}\beta_m\beta_n\beta_k\text{tr}\Big{(}\textbf{R}_{m,r}\textbf{Z}_{mk'}^H\Big{)}\text{tr}\Big{(}\textbf{R}_{n,r}\textbf{Z}_{nk'}\Big{)}\left[
p_p\sum\limits_{k''\in\mathcal{P}_{k'}}\beta_{k''}\text{tr}\left(\textbf{T}_{\omega_k}\textbf{T}_{\omega_{k''}}\right)+\sigma_v^2\text{tr}\bigg{(}\textbf{T}_{\omega_k}(\mathbf{\Gamma}_r+\mathbf{\Gamma}_t)\bigg{)}
\right].
\end{array}
\label{DL_downlink_UI}
\vspace{-8 pt}
   \end{equation}
  \vspace{-10 pt}
   \hrulefill
   \end{figure*}
Moreover, the dynamic noise caused by the active STAR-RIS is given by
    \vspace{-5pt}
 \begin{equation}
\begin{array}{ll}
\displaystyle\mathbb{E}\Big{\{}\Big{|}\textbf{g}_{k}[n]^T\boldsymbol{\Theta}_{\omega_k}^T\textbf{v}[n]\Big{|}^2\Big{\}}=\beta_k\sigma_v^2\text{tr}(S\textbf{R}\boldsymbol{\Theta}_{\omega_k}^T\boldsymbol{\Theta}_{\omega_k}^*)=\beta_k\sigma_v^2\text{tr}(\mathbf{\Gamma}_{\omega_k}).
 \end{array}
 \label{EMI_downlink}
   \end{equation}  
The analytical closed-form $\text{SINR}_k[n]$ is given in \eqref{downlink_SINR} at the top of the next page, facilitating efficient SE evaluation and providing essential insights for performance analysis and system design.
\begin{figure*}[t!]
\begin{equation}
\begin{array}{ll}
\displaystyle \text{SINR}_k[n]&\displaystyle=\frac{\displaystyle\mathbb{E}\{|\text{DS}_k[n]|^2}{\displaystyle\sum\nolimits_{k'=1}^K\mathbb{E}\{|\text{UI}_{kk'}[n]|^2\}-\mathbb{E}\{|\text{DS}_k[n]|^2+\mathbb{E}\{|\text{DN}_k[n]|^2+\mathbb{E}\{|\text{NS}_k[n]|^2}\\&\displaystyle=\frac{\displaystyle p_d\rho_k^2[n-\lambda]\Big{|}\sum\nolimits_{m=1}^{M}\sqrt{\eta_{mk}}\text{tr}\left(\textbf{Q}_{mk}\right)\Big{|}^2}{\displaystyle\sum\nolimits_{k'=1}^K\mathbb{E}\{|\text{UI}_{kk'}[n]|^2\}-p_d\rho_k^2[n-\lambda]\Big{|}\sum\nolimits_{m=1}^{M}\sqrt{\eta_{mk}}\text{tr}\left(\textbf{Q}_{mk}\right)\Big{|}^2+\beta_k\sigma_v^2\text{tr}(\mathbf{\Gamma}_{\omega_k})+\sigma^2}.
\end{array}\label{downlink_SINR}
\vspace{-5 pt}
   \end{equation}
\hrulefill
\vspace{-14 pt}
   \end{figure*}
To display the average phase error performance, when $\omega_k=\omega_{k'},~\forall k,k'$, we can define 
   \begin{equation}
\begin{array}{ll}
\displaystyle
\textbf{T}_{\omega_k}\textbf{T}_{\omega_{k'}} \displaystyle\triangleq A^4{\textbf{R}}^{1/2}\left[\bar{\textbf{R}}_{\omega_k}{\textbf{R}}\bar{\textbf{R}}_{\omega_k}-\left(\bar{\textbf{R}}_{\omega_k}{\textbf{R}}\bar{\textbf{R}}_{\omega_k}\right)\circ\textbf{I}_L
\right]{\textbf{R}}^{1/2}\\~~~~~~~~~
\displaystyle+A^4{\textbf{R}}^{1/2}\left[\left(\boldsymbol{\Phi}_{\omega_k}^H\textbf{A}^H{\textbf{R}}\bar{\textbf{R}}_{\omega_k}{\textbf{R}}\textbf{A}\boldsymbol{\Phi}_{\omega_k}\right)\circ\textbf{I}_L
\right]{\textbf{R}}^{1/2}.
\end{array}
\label{eq_22}
\end{equation}
We utilize the fractional power control to obtain power control coefficients $\eta_{mk}=\left({\sum\nolimits_{k'=1}^K{\text{tr}(\textbf{Q}_{mk'})}}\right)^{-1},~\forall k,~\forall m$ \cite{10297571,11196010}.

\vspace{-6pt}\section{Numerical Results and Discussion}
This section presents numerical results to illustrate the SE performance, emphasizing the effects of phase errors, channel aging, and amplification. The analytical results derived in \eqref{downlink_SE_description}--\eqref{eq_22} are validated by Monte Carlo (MC) simulations.

\vspace{-11pt}\subsection{Parameter Setup}
Referring to \cite{10297571}, APs are randomly placed with coordinates $x^{\text{AP}},y^{\text{AP}}\in\left[-100,100\right]$. Transmission-area users are placed with $x^{\text{user}}\in\left[450,550\right]$
and $y^{\text{user}}\in(100,150]$, reflection-area users are placed with $x^{\text{user}}\in\left[400,600\right]$
and $y^{\text{user}}\in\left[0,100\right)$ and the STAR-RIS is placed at $(x^{\text{STAR-RIS}},y^{\text{STAR-RIS}})=(500,100)$. All units of measurement for spatial coordinates are in meters.
The path loss model $\beta_x=\text{PL}_x\cdot z_x$ 
($x=mk,~m,~k$) in \cite{7827017} is utilized to acquire large-scale fading coefficients with the three-slope path loss $\text{PL}_x$ and the log-normal shadowing $z_x$. For the amplification factor matrix, $\alpha_l=\alpha, \forall l$ \cite{10025392}, $\alpha=1$ produces a passive STAR-RIS. We apply the exponential correlation model in 
\cite{qian2024impactchannelagingelectromagnetic} as the AP spatial correlation. Unless mentioned, $p_{p}=20~\text{dBm}$, $p_{d}=23~\text{dBm}$, $\sigma_v^2=-100$ dBm, $\sigma^2=-91$ dBm \cite{10264149}, %9377648
$\tau_p=5$ is the pilot sequence length, $f_c=1.9$ GHz is the carrier frequency, $T_s=0.01$ ms is the time instant length, and $d_h=d_v=\lambda/4$ for the STAR-RIS elements\cite{qian2024impactchannelagingelectromagnetic}.

\vspace{-12pt}\subsection{Results and Discussion}
\begin{figure}[!t]
	\centering	
		\centering
		\includegraphics[width=0.635\columnwidth]{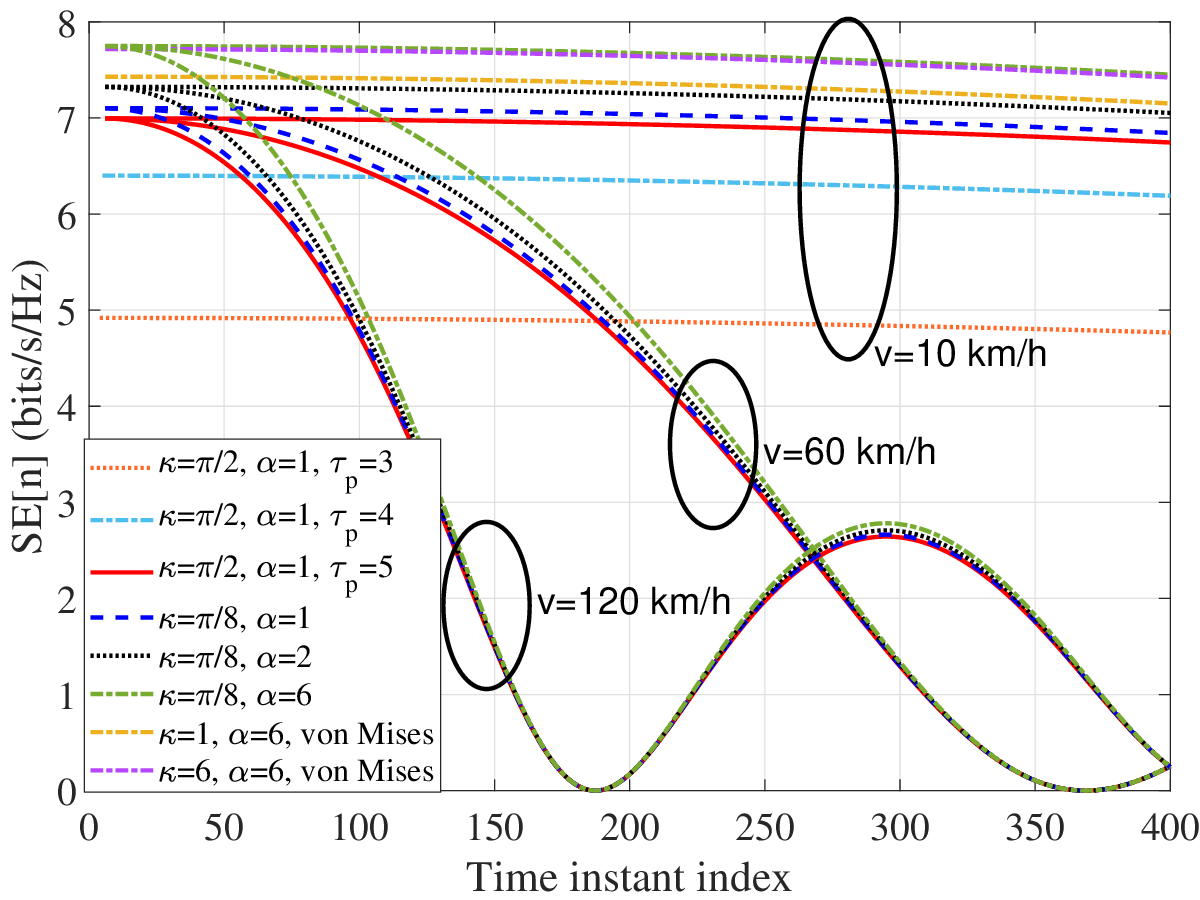}
        \vspace{-8pt}
\caption{Downlink SE versus time instant index with $M=10$, $N=4$, $K=10$, $K_t=K_r=5$, $L=64$ (Analytical Results).}
\label{fig_1}
\vspace{2 pt}
		\centering
\includegraphics[width=0.635\columnwidth]{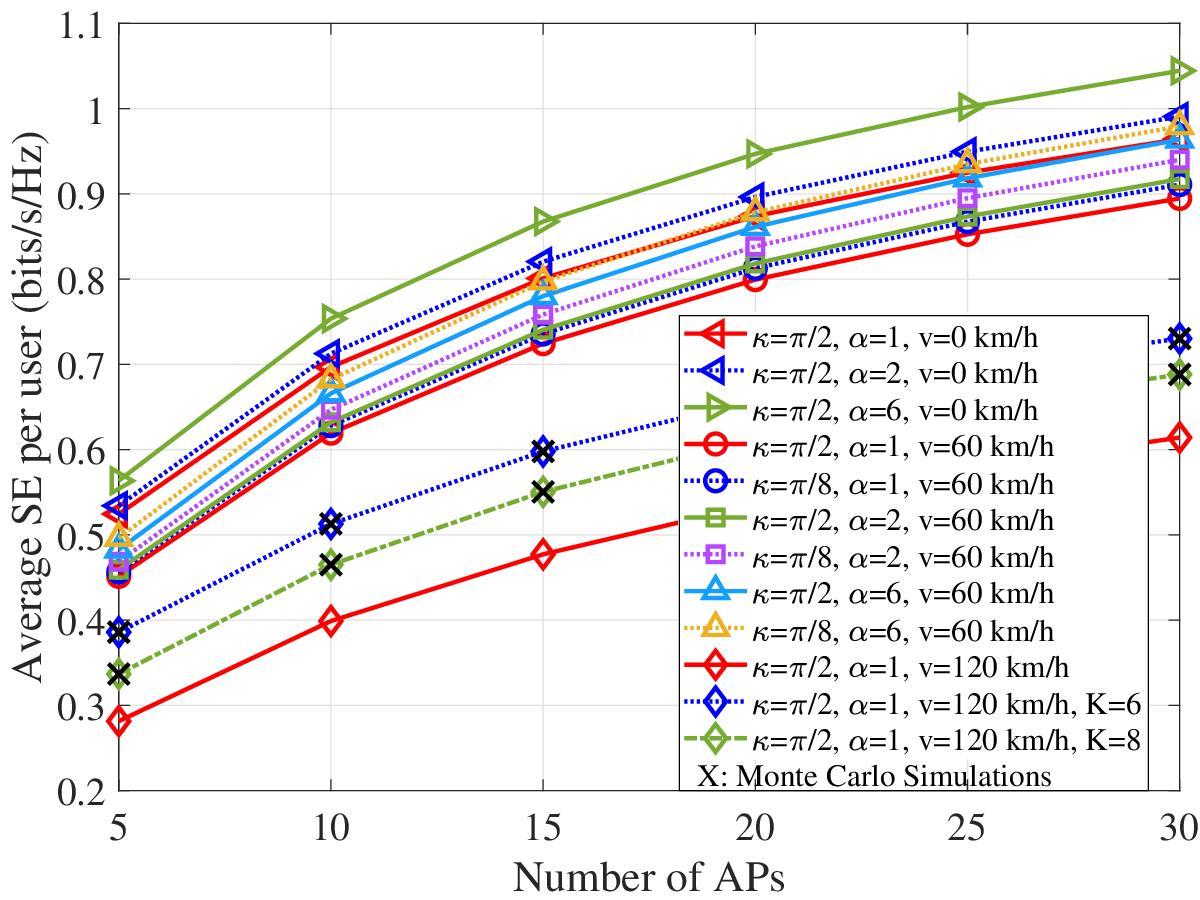}
\vspace{-8pt}
\caption{Downlink SE versus number of APs with $N=4$, $K=10$, $K_t=K_r=5$, $L=64$, $\tau_p=K/2$ (MC Simulations and Analytical Results).}
\label{fig_2}
\vspace{-20pt}
\end{figure}

Fig.~\ref{fig_1} depicts the downlink SE at the $n$-th time instant, defined as $\text{SE}[n]=\sum_{k=1}^{K}\log_2\left(1+\text{SINR}_k[n]\right)$, for the first 400 time instants, with downlink transmission starting at $n=\tau_p+1$. We observe a decreasing trend in SE as the time instant index increases, reflecting the direct impact of channel aging. This degradation is pronounced at higher user velocities, which exacerbate channel aging and cause a rapid decline in SE, shifting the first zero crossing to the left. Thus, an appropriately chosen resource block length can alleviate the impact of channel aging. As shown in Table \ref{Table_1}, the resource block length in relation to typical velocities is presented. Then, adopting a larger number of pilots, increasing from $\tau_p=3$ to $\tau_p=5$, can introduce extra channel estimation accuracy and achieve a 40\% SE gain, by introducing more pilot overhead. Notably, an active STAR-RIS with $\alpha=6$ yields a 10\% SE increase over the passive STAR-RIS at $v=10$ km/h. Increasing phase error bounds $\kappa$ consistently degrades performance; for example, increasing $\kappa$ from $\pi/8$ to $\pi/2$ leads to a $2\%$ SE reduction at $v=10$ km/h. Additionally, zero-mean von Mises distribution is adopted for comprehensive analysis, with $\mathbb{E}\{e^{i\bar{\theta}_{l}^{\omega_k}}\}=\frac{I_1(\kappa)}{I_0(\kappa)}=\phi,~\forall l,~\forall k$ \cite{9786058,10025392}, where $\kappa$ is the concentration parameter and $I_v(\cdot)$ is the modified Bessel function of the first kind with order $v$. As $\kappa$ increases from 1 to 6, phase errors become more concentrated, leading to less performance degradation. These findings highlight the need for optimizing resource block length and amplification at higher mobility to mitigate channel aging and phase error effects in practical deployments. A joint optimization tailored to varying mobility and channel conditions will be explored in future work to enhance SE by capturing parameter interdependence and balancing interference.
Thus, in subsequent analyses, all users are assigned with $v\leq 120~\text{km/h}$ and $\tau_c=182$.
\begin{table}[t!]
\caption{Resource Block Length with Different Velocities} 
\vspace{-10 pt} 
\begin{center}
\begin{tabular}{ l | c |c|c|c|c} 
\hline
{ {Velocity}} (km/h) & 60  & 90 & 120 & 150 & 180\\
\hline
\hline
$\tau_c$ (First zero position) & 364 & 244  & 182 & 146 & 122  \\  
\hline
\end{tabular}
 \end{center}\label{Table_1}
 \vspace*{-20 pt} 
\end{table}

Fig.~\ref{fig_2} presents the downlink average SE per user, $\text{SE}_\text{ave}=\frac{1}{K}\sum_{k=1}^{K}\text{SE}_{k}$, versus the number of APs $M$. The closed-form analytical results closely match the MC simulations, validating the accuracy of the theoretical analysis. Higher velocity will introduce SE degradation, reflecting the channel aging effect, as detailed in Fig. \ref{fig_1}. More users will introduce a smaller average SE akin to \cite{11196010}, due to higher inter-user interference. A larger number of APs can greatly enhance SE by providing greater spatial diversity and improved beamforming capabilities. Then, higher amplification factors at the STAR-RIS yield notable SE gains, e.g., $\alpha=2$ yields an approximate 5\% SE gain and $\alpha=6$ achieves an over 10\% SE gain compared to that of $\alpha=1$. The adverse effects of phase errors diminish with larger $M$ and $\alpha$, e.g., at $\alpha=6$, the SE loss due to severe phase errors ($\kappa=\pi/2$) reduces from 3\% at $M=5$ to 1.5\% at $M=30$, whereas at $\alpha=2$, the loss remains at 2.5\% regardless of $M$. Meanwhile, a larger $\alpha$ can mitigate the negative effects of channel aging at $v=60$ km/h to approach the stationary performance with passive or low-amplification STAR-RISs. These findings underscore the importance of jointly optimizing the number of APs and the STAR-RIS amplification factor to achieve robust performance in the presence of practical impairments such as phase errors and channel aging.

\begin{figure}[!t]
\centering
\includegraphics[width=0.635\columnwidth]{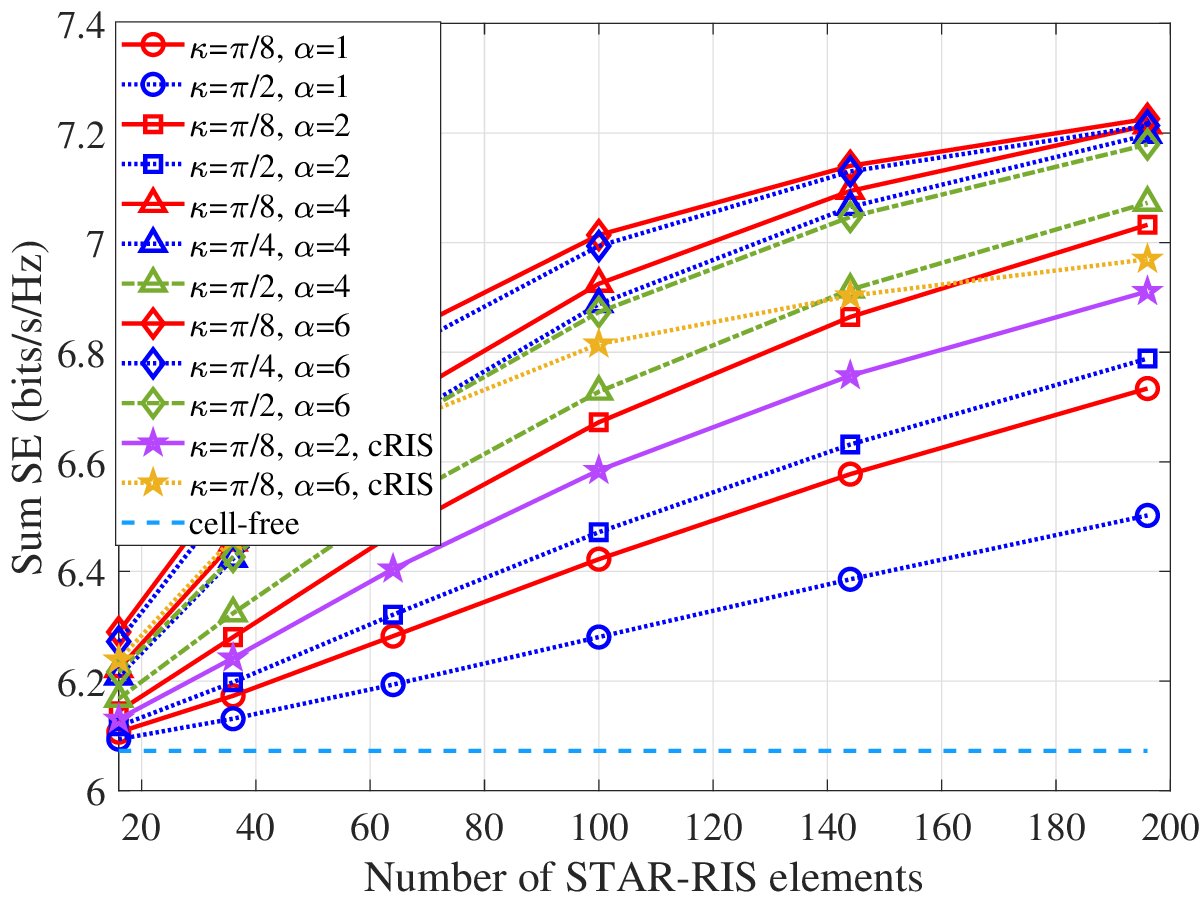}
\vspace{-6 pt}
\caption{Downlink SE versus number of STAR-RIS elements with $M=10$, $N=4$, $K=10$, $K_t=K_r=5$, $v=60$ km/h (Analytical Results).}
\label{fig_3}
\vspace{-15pt}
\end{figure}

Fig.~\ref{fig_3} illustrates the downlink sum SE versus the number of STAR-RIS elements $L$. The results indicate that increasing both the number of STAR-RIS elements and the amplification factor $\alpha$ substantially enhances SE, rendering the STAR-RIS-assisted channels more deterministic. For example, with $\kappa=\pi/2$, the SE improvement over conventional cell-free massive MIMO ranges from 1\% to 10\% for passive STAR-RIS, 1\% to 15\% for $\alpha=2$, and 4\% to 20\% for $\alpha=6$. Compared to conventional RIS (cRIS) architectures, which utilize adjacent reflecting-only and transmitting-only RISs with $L/2$ elements each, STAR-RIS-assisted systems achieve an over $4\%$ SE improvement for $L=196$. As $L$ increases, the SE loss from phase errors diminishes with higher $\alpha$: for $L=196$, $\kappa=\pi/2$ results in a 5\%, 4\%, 2\%, and 1\% SE loss at $\alpha=1$, $\alpha=2$, $\alpha=4$, and $\alpha=6$, respectively, compared to $\kappa=\pi/8$. Notably, phase errors cause greater SE degradation with a small $\alpha$ and large $L$, or with a large $\alpha$ and small $L$. The marginal benefit of increasing $\alpha$ saturates for large $L$, as amplified interference and noise may offset the gains in desired signal power. Thus, the amplification factor should be optimized rather than simply maximized in large-scale deployments to avoid excessive interference and noise. These reveal the need for joint optimization of the amplification factor and the number of STAR-RIS elements to maximize SE performance in our future work.

\vspace{-10pt}\section{Conclusion}

This work provides a comprehensive study of spatially correlated active STAR-RIS-assisted cell-free massive MIMO systems, emphasizing the impact of phase errors and channel aging. We analyzed the downlink SE performance, supported by novel closed-form SE derivations. Numerical results indicate that phase errors and channel aging greatly degrade performance. To address these effects, a resource-block-length design guideline has been proposed to counter channel aging. Also, increasing the number of APs and active STAR-RIS elements, along with larger amplification factors (e.g., $\alpha=6$), can effectively alleviate these adverse effects. 
Although active STAR-RISs with a larger $\alpha$ outperform passive STAR-RISs, conventional RISs and cell-free massive MIMO, the marginal benefit
of increasing $\alpha$ emphasizes the need to optimize rather than simply maximize $\alpha$ to balance signal gains against amplified interference and noise. 
These insights highlight the advantages of integrating active STAR-RISs into 6G networks, with future work on practical implementation challenges.
\vspace{-8 pt}
\ifCLASSOPTIONcaptionsoff
  \newpage
\fi

% if have a single appendix:
%\appendix[Proof of the Zonklar Equations]
% or
%\appendix  % for no appendix heading
% do not use \vspace{-6pt}\section anymore after \appendix, only \vspace{-6pt}\section*
% is possibly needed

% use appendices with more than one appendix
% then use \vspace{-6pt}\section to start each appendix
% you must declare a \vspace{-6pt}\section before using any
% \vspace{-11pt}\subsection or using \label (\appendices by itself
% starts a section numbered zero.)
%

%\appendices
%\vspace{-6pt}\section{Proof of the First Zonklar Equation}
%Appendix one text goes here.

% you can choose not to have a title for an appendix
% if you want by leaving the argument blank
%\vspace{-6pt}\section{}
%Appendix two text goes here.

% use section* for acknowledgment
%\vspace{-6pt}\section*{Acknowledgment}

% Can use something like this to put references on a page
% by themselves when using endfloat and the captionsoff option.
\ifCLASSOPTIONcaptionsoff
  \newpage
\fi

% trigger a \newpage just before the given reference
% number - used to balance the columns on the last page
% adjust value as needed - may need to be readjusted if
% the document is modified later
%\IEEEtriggeratref{8}
% The "triggered" command can be changed if desired:
%\IEEEtriggercmd{\enlargethispage{-5in}}

% references section

% can use a bibliography generated by BibTeX as a .bbl file
% BibTeX documentation can be easily obtained at:
% http://mirror.ctan.org/biblio/bibtex/contrib/doc/
% The IEEEtran BibTeX style support page is at:
% http://www.michaelshell.org/tex/ieeetran/bibtex/
\bibliographystyle{IEEEtran}
% argument is your BibTeX string definitions and bibliography database(s)
\bibliography{IEEEabrv,QIAN_WCL2025_2454_ref.bib}
\end{document}